\documentclass[11pt,letter]{article}
  \usepackage{graphicx}
  \usepackage{epstopdf}
  \usepackage{color}
  \usepackage{epsfig,amssymb,amsmath}
  mm\evensidemargin=-5true mm \topmargin=-15true mm

\usepackage[
      colorlinks=true,
      linkcolor=blue,
      urlcolor=blue,
      filecolor=blue,
      citecolor=red,
      pdfstartview=FitV,
      pdftitle={},
      pdfauthor={},
      pdfsubject={},
      pdfkeywords={},
      pdfpagemode=None,
      bookmarksopen=true
]{hyperref}

\definecolor{geo}{rgb}{0,0.0,0.65}
\definecolor{GC}{rgb}{0,0.0,0.65}

  \newcommand{\be}{\begin{equation}}
  \newcommand{\ee}{\end{equation}}
  \newcommand{\bea}{\begin{eqnarray}}
  \newcommand{\eea}{\end{eqnarray}}

\def\cL{\mathcal{L}}

\newcommand{\beq}{\begin{equation}}
\newcommand{\eeq}{\end{equation}}
\newcommand{\beqs}{\begin{eqnarray}}
\newcommand{\eeqs}{\end{eqnarray}}

\date{}

\begin{document}
\title{No-hair theorems for a static and stationary reflecting star}
\author{Srijit Bhattacharjee\footnote{srijitb@iiita.ac.in} \\ Indian Institute of Information Technology, Allahabad \\Devghat, Jhalwa, Uttar Pradesh-211012, India.\\
\\ Sudipta Sarkar\footnote{sudiptas@iitgn.ac.in}\\ Indian Institute of Technology, Gandhinagar, Gujarat, 382355, India.}

\maketitle
\abstract{We prove the non existence of massive scalar, vector and tensor hairs outside the surface of a static and stationary compact reflecting star. Our result is the extension of the no hair theorem for black holes to horizonless compact configurations with reflecting boundary condition at the surface. We also generalize the proof for spacetimes with a positive cosmological constant.}
\section{Introduction}
Over the last century, Einstein's ideas have taught us how gravity affects the structure of spacetime, sometimes in a very dramatic way, leading to regions that are causally inaccessible to any observer. The prototypical example of this is the event horizon of a black hole, the surface of no return for any in-falling object which is formally defined as the boundary of the region from which signals cannot reach null infinity. In general relativity (GR), one of the most intriguing properties of a black hole is the no hair theorem, which asserts that an asymptotically flat, regular black hole can not support self-gravitating static matter configurations made of massive scalar, vector and tensor fields in their external spacetime regions \cite{Beken-PRL, Beken-static,Beken-station}.\\

There has been a plethora of investigations performed to find examples where the no hair theorem doesn't work. In fact for many Einstein-matter systems, it has been shown that the no hair theorem can be evaded. See, e.g., Ref. \cite{Bekenstein:1996pn} and references therein. For a more recent and comprehensive review, see Ref. \cite{Cardoso:2016ryw}. However, most of these solutions either have been shown to be unstable or they are obtained by relaxing one of the assumptions considered by Bekenstein \cite{Cardoso:2016ryw, Herdeiro:2015waa, Hod:2012px}. \\

The no hair theorems originate from the existence of the static event horizon and may also provide crucial observational input to distinguish black holes from other possible configurations. Therefore, an important question is to ask whether these no hair theorems are the unique features of black objects or similar uniqueness results also hold for non black hole configurations of General Relativity.\\ 

Recently, Hod \cite{Hod:2016vkt} presented a version of no hair theorem for spherically symmetric compact objects showing that such configurations cannot support static scalar fields with reflecting boundary conditions on the surface. The theorem immediately rules out the existence of massive scalar hair outside the surface of a spherically symmetric compact reflecting star. This result is interesting because this is the first illustration of the no hair type theorem for horizonless configurations. Operationally, the theorem shows that the ingoing absorbing boundary condition on the matter fields at the event horizon can be replaced by a reflecting boundary condition on the surface of the star without any modification of the characteristic no-scalar-hair property with asymptotically flat boundary conditions.\\

 Although such an analysis may not have direct application to any real astrophysical object but it is useful to study this more to understand the mathematical structure of horizonless compact objects. It is worth mentioning here that in the `black-hole bomb' scenario introduced by Press and Teukolsky {\cite{Press-Teukolsky}, one places a reflecting mirror around a black hole to prevent the scalar field from escaping to infinity. Here, if one wants to prevent the scalar field outside the star from entering the center of it then this reflecting boundary condition is the most natural choice. Also, examining the no-hair conjecture in this simple setting can provide a way to study similar features of horizonless objects like boson stars or gravastars. But, the main motivation is to comprehend the scope of applicability of no hair type theorems in general relativity. As a result, Hod's result demands more detailed examination. In particular, the original proof assumes spherical symmetry and only uses scalar matter. In this work, we generalize the result by Hod \cite{Hod:2016vkt} beyond spherical symmetry and also for massive Proca and massive spin two tensor fields. We show that, in all cases, the reflecting boundary condition is indeed sufficient to prove the non existence of solutions. We also discuss the extension of this result to spacetimes with positive cosmological constant. Our result implies that the no hair theorem for compact stars can be made as rigorous as the standard no hair theorems for black holes.

\section{Scalar field}
Let us assume that a compact star is coupled to a set of real scalar fields $\phi_k$. We describe these fields in terms of a {\it canonical}\footnote{By canonical Lagrangian we mean the fields possess canonical kinetic terms and are minimally coupled to gravity.} Lagrangian density $\cal{L}$. After varying the Lagrangian with respect to $\phi$ we get the field equation:

\be
\sqrt{-g}\left[\sqrt{-g}\frac{\partial \cal{L}}{\partial \phi_{k,\mu}}\right]_{,\mu} -\frac{\partial \cal{L}}{\partial\phi_{k}}= 0
.\label{eqm}\ee

Next, we integrate  Eq.(\ref{eqm}) over the exterior region of the star after multiplying it by $\sqrt{-g}\, \phi_k\, d^4x$, and convert one of the terms into a surface integral. Finally summing over the fields we get \cite{Beken-static},
 \be
- \int b^{\mu} dS_{\mu} + \sum_k \int \left(\phi_{k,\mu}\frac{\partial \cal{L}}{\phi_{k, \mu}} +\phi_{k}\frac{\partial \cal{L}}{\partial\phi_{k}}\right)\sqrt{-g}\,d^4x=0,
 \label{eq2}\ee
where we have defined,
\[b^{\mu}=\sum_k \phi_k\frac{\partial \cal{L}}{\phi_{k, \mu}}.\]
The area segment $dS_{\mu}$ is to be taken at various boundaries of the space time (both space like and time like infinities) and also on the surface of the star. We now evaluate the surface terms on various such boundaries of the space time. Note, that for every physically relevant field the quantity $b^{\mu}$ vanishes asymptotically as $1/r^{3}$ for the massless case, and exponentially for the massive case. Thus there are no contributions from spatial infinity to the boundary integral in Eq.(\ref{eq2}). For the time like boundary, we can always arrange the normal vector in a way such that it doesn't have any spatial component, i.e. we have $n_i=dS_i=0$. We assume that the field shares the symmetry of the background space time. Then $b^0=0$ because of time independence and this leads to $b^{\mu}dS_{\mu}=0$ at the time like boundary as well. Now we are left with the reflecting surface of the star. Note that $b^{\mu}b_{\mu}$ is a physical scalar everywhere including the surface of the star as it can be expressed in terms of invariants constructed out of stress-energy tensor $T_{\mu\nu}$. Now, we impose a reflecting boundary condition on the scalar field at the surface of the star implying $\phi_k(\textrm{surface})=0$. This ensures that $b^{\mu}$ vanishes identically at the surface of the star. Note that, if we were considering the black hole horizon case, the horizon boundary condition would have ensured the same. The reflecting boundary condition at the surface of the star also leads to the same result. Hence entire surface contribution of Eq.(\ref{eq2}) is zero and we have,
 
 \be 
\sum_k \int \left(\phi_{k,\mu}\frac{\partial \cal{L}}{\phi_{k, \mu}}+ \phi_{k}\frac{\partial \cal{L}}{\partial\phi_{k}}\right)\sqrt{-g}\,d^4x=0.
\label{condn}\ee

For a canonical scalar field the Lagrangian density reads:
\be \cL= -{1 \over 2} \left(\partial_{\alpha} \phi\,\partial^{\alpha} \phi + V(\phi)\right).\ee

The potential is assumed to be convex and non-negative and therefore the second term in Eq.(\ref{condn}), $\phi\, V_{,\phi}$ is positive definite throughout the domain of our interest. We have assumed that the matter also obeys staticity and therefore Killing vector $\xi^\mu$ that generates time symmetry should satisfy  $<\xi,d\phi>={\cal L}_{\xi}\phi=0$. As a result, $\partial_\mu \phi$ is nowhere time like in the exterior of the star. Therefore, the first term in Eq.(\ref{condn}), which for a canonical Lagrangian is of the form $(\partial\phi)^2$ is also non-negative \cite{Heusler:1996ft}. Hence the only possible way the integral in Eq.(\ref{condn}) can vanish is to have the scalar field identically zero throughout the region of interest. This immediately implies the non existence of any nontrivial solution. This proof is also valid for a massless ($m=0$) scalar field for which the potential $V$ may be zero. In that case we need the scalar field $\phi$ to be a constant throughout the exterior region, but as we have assumed the scalar field is vanishing in the asymptotic regions, we can only accommodate the $\phi=0$ solution everywhere. This proof disqualifies a scalar hair for a reflecting star whose domain of outer communication is {\it strictly stationary} i.e. configurations without an ergoregion. \\

This proof can be easily generalized for stationary axisymmetric spacetimes (rotating) \cite{Beken-station}. In this case let there be a canonical scalar field $\psi$ that possesses symmetries of the spacetime at equilibrium: $\partial_t \psi=\partial_{\phi}\psi=0$. Then using similar steps as in the static case we again arrive at Eq.(\ref{eq2}). \footnote{One can always find an orthonormal frame in a stationary spacetime and evaluate all tensor components in that frame. If one does that the proof for stationary case becomes almost identical to the one depicted for static case. However, we don't resort to any particular frame to show this proof here.} Now, for the outer boundaries we discard any contribution from the first term of this equation. For inner boundary i.e. on the star's surface we again use the reflecting boundary condition to get rid of the $b^{\mu}dS_{\mu}$ term as $b^{\mu}$ will again be proportional to the scalar field itself. Thus for a stationary spacetime with $t-\phi$ isometry, we again don't have any nontrivial scalar field in the exterior of the star.

\section{Proca field}
In this section we consider the case of a massive vector field sourcing a static spacetime. A no-hair result for such fields has already been established for asymptotically flat black holes with a regular event horizon. We now want to generalize the same for the case of a reflecting compact object. In this case the Lagrangian reads,
\be
\cL=-\frac{1}{4}\left(F_{\mu\nu}F^{\mu\nu}\,+\,m^2B^{\mu}B_{\mu}\right)
.\ee
The Proca field $B^{\mu}$ is a physical field in the sense that it doesn't transform under a gauge transformation and therefore is free from all gauge ambiguities. The field strength is $F_{\mu\nu}=B_{\mu,\nu}-B_{\nu,\mu}$, and the equation of motion is given by,

\be
F^{\mu\nu}{;\nu}\,+m^2B^{\mu}=0.
\label{proca}\ee
The Proca field equation should be invariant under time reversal even when a source term is present. The time reversal symmetry in the matter sector demands that $B_i$ and $F^{ij}$ must vanish in the static case. Next, retracing the same steps as in the case of scalar fields we obtain,

\be
- \int b^{\mu} dS_{\mu}\,+\,\int g_{00}\left[g_{ij}F^{0i}F^{0j}\,+\,m^2(B^0)^2\right]\sqrt{-g}\,d^4x=0,
\label{procacnd}\ee

where the boundary term is $b^{\mu}=-F^{\mu\nu}B_{\nu}$. Since $b^0$ will again be zero in the static case and on the surface of the star due to the reflecting boundary condition $B^{\mu}=0$, we again have no contribution from the boundary terms and are only left with the integral in the bulk in Eq. (\ref{procacnd}),

\be
\int g_{00}\left[g_{ij}F^{0i}F^{0j}\,+\,m^2(B^0)^2\right]\sqrt{-g}\, d^4x=0.
\label{intgrnd}\ee

Note that this again assumes the regularity of the derivatives of the field on the surface of the star. Outside the reflecting star the spacetime is static and $g_{ij}$ is positive definite \cite{Beken-static}. The argument considers a static spacetime and the fact that $g_{00}$ can never become positive in the outer region of a black hole. Note that $g_{00}$ is the norm of the Killing vector that generates time translation in a static spacetime. Moreover $g_{00}$ reaches $-1$ at asymptotic infinity. Since our spacetime is static and given the fact that any $g_{00}=0$ surface in a static spacetime must be null \cite{Vishu} we discard the possibility that $g_{00}$ becomes positive in the exterior region of the star. Hence $g_{ij}$ is a positive definite matrix and the integrand in Eq. (\ref{intgrnd}) is negative definite. Therefore the integrand can only be zero if the Proca field itself vanishes everywhere. So, we conclude that a massive neutral vector field is not supported by a perfectly reflecting star. \\

For a massless field the above conclusion cannot be made. In the massless case, the $B_{\mu}$ field will allow gauge transformations like $B_{\mu} \rightarrow B_{\mu} \,+\,\partial_{\mu}\psi(x)$ and the boundary term $b^{\mu}=-F^{\mu\nu}B_{\nu}$ can no longer be regarded as physical. Hence the inferences that have been made for the massive case related to the boundary integrals of Eq. (\ref{condn}) now become gauge dependent. Therefore we cannot discard the possibility of having a perfectly reflecting star with a massless vector field in this setup.

\section{Spin-2 tensor field}
We start with the following Lagrangian for a massive spin- $2$ tensor field $h_{\mu \nu}$,

\be 
\cL=-\frac{1}{2}\left( h_{\mu\nu ;\alpha}h^{\mu\nu; \alpha}+m^2h_{\mu\nu}h^{\mu\nu}\right)
.\ee

In a static spacetime, the massive graviton field satisfies $h_{\mu\nu,0} =0$. We again proceed in a similar way as we did for the scalar field and obtain an equivalent condition like Eq. (\ref{eq2}). Here the boundary term becomes,

\be
b^{\mu}=-h_{\alpha\beta}h^{\alpha\beta;\mu}
.\ee

Next, we consider the fact that for static case, $b^0=0$ and the normal to the surface at time like infinity can again be chosen so as to have only time component. Hence the boundary integral at time like infinity vanishes. Also as the field $h_{\mu\nu}$ falls off exponentially as it approaches spatial infinity we don't have any contribution from the boundary term at spatial infinity. Further due to the reflecting boundary condition $h_{\mu\nu}=0$ at the surface, (for all non-zero components of massive spin-2 field) and one obtains that the contribution from the surface of the star also zero. Hence we are left with the following integral,

 \be 
 \int \left(h_{\mu\nu;\alpha}h^{\mu\nu;\alpha}+ m^2\,h_{\mu\nu}h^{\mu\nu}\right)\sqrt{-g}\,d^4x=0
\label{condn1}.\ee

Since the space time is static and we also demand that the matter field shares the symmetry of the spacetime, we find

\[h_{\mu\nu,0}=0\qquad h_{0i ;j}=0.\]

Also, $g_{ij}$ is a positive definite metric and therefore the first term in Eq. (\ref{condn1}) is positive definite. Further, one can show by using an orthonormal frame that at any arbitrary point outside the star the metric $g_{ij}$ can be diagonalized and the eigenvalues $\lambda_i$ are positive definite \cite{Beken-station}. We thus have,
\be
h_{\mu\nu}h^{\mu\nu}=(g_{00}h^{00})^2\,+\,\sum_{ij}\lambda_i\lambda_j(h^{ij})^2 \geq 0.
\ee

The kinetic term in Eq. (\ref{condn1}) can also be shown to be positive semi-definite by the above construction. Therefore the only way the condition Eq. (\ref{condn1}) can hold is to have the massive spin-2 field trivially zero everywhere outside the star.

By the same reasoning as been made in the massless spin-1 case we cannot discard the possibility of having a massless spin-2 field supported by spacetime outside a reflecting star.  

\section{Spacetime with $\Lambda > 0$}

The no-hair conjecture for static reflecting star can also be proven for a space time with a positive cosmological constant $\Lambda>0$. For this, we assume a spherically symmetric space time outside a reflecting star in the presence of positive cosmological constant. We concentrate on the part of spacetime between the cosmological horizon and the surface of the star. The norm of the Killing vector $\xi$, $\lambda(r)=\sqrt{-\xi.\xi}$ that generates the time translation symmetry vanishes at the cosmological horizon. We take a space like hyper surface $\Sigma$ between the star and cosmological horizon and prove the no-hair theorem. We use a projector $\Pi_{\mu}^{\nu}=\delta^{\nu}_{\mu}\,+\,\lambda^{-2}\xi^{\nu}\xi_{\mu}$ to project all spacetime vectors on the hyper surface. With the aid of the projector one can use the following identity for any rank $p$ antisymmetric tensor $\Omega$ whose Lie derivative with respect to $\xi^{\mu}$ vanishes \cite{Bhattacharya:2007zzb}:

\be
\tilde{\nabla}_{\alpha}(\lambda\omega^{\alpha\mu\nu\cdots})=\lambda(\nabla_{\alpha}\Omega^{{\alpha'\mu'\nu'\cdots}})\Pi^{\mu}_{\mu'}\Pi^{\nu}_{\nu'}\cdots
,\ee

where $\tilde{\nabla}$ is the induced connection on $\Sigma$ and $\omega$ is the projection of $\Omega$ on $\Sigma$ . For a scalar field with a convex potential $V(\phi)$, we may project the equation of motion using above identity and get:
\be
\tilde{\nabla}_{\alpha}(\lambda\tilde{\nabla^{\alpha}}\phi)=\lambda V'(\phi)
.\ee
We now multiply both sides of this with $V'(\phi)$ and integrate to get
\be
\int_{\partial\Sigma} \lambda V'(\phi)n^{\mu}\tilde{\nabla}_{\mu}\phi\,+\,\int_{\Sigma} \lambda[V''(\phi)\tilde{\nabla}^{\mu}\phi\tilde{\nabla}_{\mu}\phi\,+\,V'(\phi)^2]=0\label{dscondn}
.\ee
The boundary integral now consists of two regions: a cross section of the cosmological horizon and that of the surface of the star. On the cosmological horizon $\lambda \rightarrow 0$, as a result the contribution to the boundary term vanishes trivially. For the inner boundary (i.e. on the surface of the star) if $V'(\phi)$ is zero there is no contribution. For polynomial type potentials (including a canonical massive scalar field) this is satisfied upon invoking the Dirichlet condition on the field. So for this class of potentials, we can set the boundary term on the inner boundary also to be zero. The bulk term in Eq. (\ref{dscondn}) is positive definite as $\tilde{\nabla}^{\mu}\phi\tilde{\nabla}_{\mu}\phi$ is positive on the space like surface $\Sigma$ and for the convexity of the potential: $V''(\phi)>0$. Therefore the bulk term can only vanish if $\phi$ is zero everywhere. Hence the no-hair result. 

This no hair result can also be generalized for a stationary axisymmetric spacetime having a reflecting star. The proof proceeds in a similar way as in the scalar case \cite{Bhattacharya:2011dq}.

\section{Discussions}

The validity of no hair type theorems for matter fields outside reflecting stars shows that horizonless compact reflecting stars also share the no-hair property similar to the regular asymptotically flat black holes. Mathematically, the key point is the parallel between the regularity condition at the horizon of black holes and the Dirichlet boundary condition on the surface of the star.  Our work extends the no-hair result beyond the spherically symmetric setting and proves the same for various matter fields in static and stationary configurations. This definitely establishes the generality of the result obtained in Ref. \cite{Hod:2016vkt}.\\

A future direction could be to understand if such a no hair theorem can be proven for non-canonical scalar fields as in the case of black holes. Unlike the proofs presented here, black hole no hair theorem for non canonical scalar fields requires the use of Einstein's field equations. As a result, it would be interesting if such a result can also be extended to reflecting starts.\\

No-hair theorems were considered to be a unique feature of horizons in general relativity. The uniqueness of black hole solutions with asymptotically flat boundary conditions is also a manifestation of the no hair property of black objects. Since the black hole event horizon does not have any hair apart from the mass, angular momentum and charge, the process of gravitational collapse leads to the decay of higher multipoles \cite{Price} and the collapse ends in an unique static/stationary black hole state \cite{Chrusciel:2012jk}. The extension of the no hair theorem for reflecting stars opens up an interesting possibility of deriving similar decay laws for collapsing matter configurations forming a reflecting star.\\

Another extension of our work could be to establish a no hair (scalar) theorem for a reflecting star by relaxing one of the assumptions of the proof. For example, a possible investigation could be to prove the theorem with a time dependent scalar field outside the reflecting star \cite{Graham:2014ina}. There are examples of hairy black holes where the scalar field does not inherit the symmetry of the spacetime but recent studies reveal that the possibility of such solutions are highly constrained \cite{Smolic:2015txa, Smolic:2016dmh}. Therefore it would be interesting to find such constraints in the case of horizonless objects. One could also explore the possibility of a scalar hair for a reflecting star or any horizonless compact object violating one of the energy conditions. Analyzing the situation with an anti De Sitter asymptotic boundary condition is also an interesting possibility.

\section{Acknowledgements}

SS is supported by the Department of Science and Technology, Government of India under the SERB Fast Track Scheme for Young Scientists (YSS/2015/001346).

\end{document}